\documentclass[11pt]{article}

\makeatletter

\@addtoreset{equation}{section}
\makeatother
\usepackage{amsfonts}
\usepackage{eucal}
\usepackage{amsmath}

\newtheorem{theorem}{\bf Theorem}[section]

\newtheorem{corollary}[theorem]{\bf Corollary}

\newtheorem{remark}[theorem]{\bf Remark}
\newtheorem{example}[theorem]{\bf Example}

\newcommand{\lang}{\left\langle}
\newcommand{\rang}{\right\rangle}
\renewcommand{\Re}{{\rm Re}\,}

\newcommand{\qed}{\hbox{\rule{4pt}{7pt}}}

\newcommand{\MSH}{{\cal H}}

\begin{document}
\title{\bf 
Convergence Conditions 
of Mixed States and their von Neumann Entropy \\
 in Continuous Quantum Measurements
}
  \author{Toru Fuda\footnote{E-mail:
t-fuda@math.sci.hokudai.ac.jp}\\
Department of Mathematics\\
Hokkaido University\\
Sapporo 060-0810\\
Japan
}
\date{}

\maketitle

\begin{abstract}
By carrying out appropriate continuous quantum measurements with
 a family of projection operators, a unitary channel can be 
approximated in an arbitrary precision in the trace norm sense. 
In particular, the quantum Zeno effect is described as an application. 
In the case of an infinite dimension, although the von Neumann entropy 
is not necessarily continuous, the difference of the entropies between 
the states, as mentioned above, can be arbitrarily made small under 
some conditions.
\end{abstract}

\medskip

\section{Introduction}

The quantum Zeno effect (QZE) is a quantum effect which 
was shown by Misra and Sudarshan in \cite{MS}.
This effect demonstrates that, in quantum mechanics, 
continuous measurements can freeze a state.
Of course, this effect is peculiar to quantum mechanics.
Such an effect is not observed in classical mechanics. 
The QZE has been extensively investigated 
by many researchers since its discovery. 

Recently, some general mathematical aspects of quantum Zeno effect 
were investigated in \cite{AF}. 
In particular, continuous measurements of 
a state along a certain curve in a Hilbert space were considered.
Roughly speaking, continuous measurements made along a 
curve prescribed in advance change the initial state to 
the final state with probability $1$. 
This fact includes the QZE as a special case. 
However, in the paper \cite{AF}, it is assumed that states 
under consideration are vector states. 

In this paper, we show that a result similar to one in \cite{AF} 
holds with respect to mixed states too. 
By considering a mixed state, its von Neumann entropy can 
also be considered. 
In the case where the Hilbert space under consideration
is infinite dimensional, the von Neumann entropy is not 
necessarily continuous with respect to the trace norm.
Hence, by continuous measurements, even if the initial state 
converges to the final state in the trace norm sense,
it does not always mean that the von Neumann entropy converges too.
Moreover, the set of density operators with finite entropy 
is a first category \cite{We}.
Hence, it is meaningful to investigate convergence conditions
of the von Neumann entropy in our continuous measurements.

In Section 2, we begin with defining the 
``continuous quantum measurements" as a certain type of 
quantum channel. 
We use two types of quantum channels and a combination of them.
By doing so, a concept of ``continuous quantum measurements" 
are defined clearly.
We consider conditions for pointwise convergence 
and trace norm convergence. 
We apply obtained results to the QZE.

In Section 3, we consider the von Neumann entropy in infinite 
dimension. We show that the convergence conditions of the 
von Neumann entropy in continuous quantum measurement which 
considered in Section 2.
Here, Simon's convergence theorem \cite{LR} plays a central role.


\section{Continuous measurements for mixed states}

\subsection{Preliminaries}

Let $\mathcal{H}$ be a separable Hilbert space 
of state vectors of a quantum system $\mathcal{S}$.
We denote the inner product and the norm of $\MSH$ by
$\lang \,\cdot\,,\,\cdot\,\rang$ 
(anti-linear in the first variable and linear in the second) 
and $\|\,\cdot\,\|$, respectively.
Let $d (\leq \infty)$ be the dimension of $\mathcal{H}$.
We denote all bounded linear operators, 
all compact operators, all trace-class operators, 
all density operators, and all unitary operators on $\mathcal{H}$ 
by $\mathfrak{B}(\mathcal{H}), \mathfrak{C}(\mathcal{H}), 
\mathfrak{T}(\mathcal{H}), \mathfrak{S}(\mathcal{H}), and 
\ \mathfrak{U}(\mathcal{H})$, respectively.
A mixed state of $\mathcal{S}$ is represented as an element of 
$\mathfrak{S}(\mathcal{H})$.
We denote the trace norm by $\|\cdot \|_1:=\mathrm{Tr}|\cdot|$.
The Hamiltonian of the quantum system $\mathcal{S}$ is given 
by a self-adjoint operator $H$ which is time independent.
The domain of $H$ is denoted as $D(H)$.

Let us consider the following two maps on $\mathfrak{S}(\mathcal{H})$:

\begin{enumerate}
	\item {(\textbf{Unitary channel})}

	Let $U$ be a unitary operator on $\mathcal{H}$ and 
	$\mathcal{E}_U$ be a map on $\mathfrak{S}(\mathcal{H})$ 
	which is given by
	$$
		\mathcal{E}_U\rho := U\rho U^*, 
		\ \forall \rho \in \mathfrak{S}(\mathcal{H}).
	$$
    In particular, in the case 
	$U=e^{-itH}\ (t\in \mathbb{R})$,
	we denote $\mathcal{E}_{e^{-itH}}$ by $\mathcal{E}_t$.

	\item {(\textbf{Projection channel})}

	Let $\mathfrak{P}:=\{P_n\}_n$ be a family of projection 
	operators
	on $\mathcal{H}$ with $P_m\perp P_n\ (m \neq n), I=\sum_nP_n$, 
	and $\mathcal{E}_{\mathfrak{P}}$ be a map on 
	$\mathfrak{S}(\mathcal{H})$ which is given by
	$$
		\mathcal{E}_{\mathfrak{P}}\rho := \sum_n P_n\rho P_n, 
		\ \forall \rho \in \mathfrak{S}(\mathcal{H}).
	$$
\end{enumerate}

Now, consider a state $\rho \in \mathfrak{S}(\mathcal{H})$ fixed
and suppose that one of the Schatten decompositions is given by
\begin{equation}
	\rho 
	= \sum_{n=1}^d \lambda _n|\Psi_n \rangle \langle \Psi_n|,
		\label{schatten}
\end{equation}
where, for all $\Psi , \Phi \in \mathcal{H}$, 
we denote the operator $\langle \Psi , \cdot \ \rangle \Phi$ by
$| \Phi \rangle \langle \Psi |$.
In (\ref{schatten}), 
we allow $\lambda_n = 0$ to take $\Psi_n$ such that 
$\{ \Psi_n\}_{n=1}^d$ is a complete orthonormal system (CONS).
We remark that it is not necessarily 
$\lambda_n \geq \lambda_{n+1}$ in this representation.

Let us consider a time interval $[0,\tau]$ with $\tau>0$.
For the decomposition (\ref{schatten}), consider a CONS 
of $\mathcal{H}$ denoted by $\{\Psi_n(t)\}_{n=1}^d$ 
which is parametrized by 
$t\in [0, \tau]$ with $\Psi_n(0)=\Psi_n \ (1\leq \forall n \leq d)$.
If $n\in \mathbb{N}$ is fixed, then 
$\Psi_n (\cdot)$ is a map from $[0, \tau]$ to $\mathcal{H}$.

We define 
\begin{equation}
	\mathfrak{P}(t)
	:=\{|\Psi_n (t) \rangle \langle \Psi_n (t)| \}_{n=1}^d,
	\quad (t\in [0, \tau]).
\end{equation}

Let $\Delta: t_0, t_1,\cdots, t_N$ ($t_j \in [0,\tau], j=0,\cdots,N$) 
be an arbitrary  partition  of the  interval $[0,\tau]$:
$$
	0=t_0 < t_1 < \cdots < t_{N-1} < t_N = \tau.
$$

We set
$$
	\Delta _k :=t_k -t_{k-1}, \quad (k=1,\cdots ,N), 
	\qquad |\Delta| :=\max_{1 \leq k \leq N}\Delta _k,
$$
and define
\begin{eqnarray}
	\rho_{\Delta}(\tau)
	:=\mathcal{E}_{\mathfrak{P}(t_N)} \circ 
	\mathcal{E}_{\Delta_N}\circ 
	\mathcal{E}_{\mathfrak{P}(t_{N-1})} \circ 
	\mathcal{E}_{\Delta_{N-1}} \circ 
	\cdots \circ \mathcal{E}_{\mathfrak{P}(t_1)}\circ
	\mathcal{E}_{\Delta_1}\rho.
\end{eqnarray}
In the context of quantum mechanics where $\rho_{\Delta}(\tau)$
is interpreted as the posterior state that, 
in the successive measurements at time
$t_1 , \cdots , t_N$ by using the family of projection operators 
$\mathfrak{P}(t_1), \cdots , \mathfrak{P}(t_N)$, respectively.
We remark that $\rho_{\Delta}(\tau)$ is dependent on 
the form of decomposition (\ref{schatten}).

If $\rho_{\Delta}(\tau)$ converges 
with respect to $|\Delta|\to 0$ in a certain sense, 
we call such a measurements of a series ``continuous quantum
measurements".

By direct computations, we have
\begin{equation}
	\rho_{\Delta}(\tau) = \sum_{k}\lambda_{\Delta, k}
	\left|\Psi_k(\tau) \rangle \langle \Psi_k(\tau)\right|
\end{equation}
with
\begin{eqnarray}
	\lambda_{\Delta, k}
	:=\sum_{k_0,\cdots,k_{N-1}}
	\lambda_{k_0}\prod _{j=1}^N\left|\langle 
	\Psi_{k_j}(t_j), e^{-i\Delta_j H} 
	\Psi_{k_{j-1}}(t_{j-1})\rangle \right|^2, \quad (k_N = k).
\end{eqnarray}

\subsection{Pointwise convergence}

Let us consider a convergence condition of $\lambda_{\Delta, k}$
in the case $|\Delta |\to 0$.

Let
\begin{eqnarray}
	\gamma_{\Delta, k}
	&:=&\prod_{j=1}^{N}\left|\langle \Psi_{k}(t_j), e^{-i\Delta_j H} 
	\Psi_{k}(t_{j-1})\rangle \right|^2 ,\\
	\epsilon_{\Delta, k}
	&:=&\sum_{\stackrel{k_0, \cdots, k_{N-1}}
	{\exists l\in\{0,\cdots,N-1\}, k_l\neq k}}
	\lambda_{k_0}\prod _{j=1}^N
	\left|\langle \Psi_{k_j}(t_j), e^{-i\Delta_j H} 
	\Psi_{k_{j-1}}(t_{j-1})\rangle \right|^2,
\end{eqnarray}
so that
\begin{equation}
	\lambda_{\Delta, k} = \lambda_k \gamma_{\Delta, k}
	+\epsilon_{\Delta, k}.\label{decomep}
\end{equation}

\medskip

\begin{theorem}\label{saisho}
	Assume that there exists $k\in \mathbb{N}$ 
	such that the following conditions hold:
	\begin{eqnarray}
		&&\forall \lambda \in [0, \tau],\quad \Psi_k(\lambda) 
		\in D(H), \label{A0} \\
		&&\xi_k 
		:= \sup_{0\leq \lambda\leq \tau}
		\|H\Psi_k(\lambda)\|<\infty, \label{A1} \\
		&&\eta_k :=\sup_{\stackrel{\lambda, \nu \in [0, \tau]}
		{\lambda \neq \nu}}\frac{\|\Psi_k(\lambda)-\Psi_k(\nu)\|}
		{|\lambda - \nu|}<\infty,\label{A2}\\
		&&\lim_{|\Delta|\to 0}\sum_{j=1}^N
		\Re \lang \Psi_k(t_j)-\Psi_k(t_{j-1}), \Psi_k(t_{j-1})\rang 
		= 0. \label{A3}
	\end{eqnarray}
	Then we have 
	\begin{equation}
		\lim_{|\Delta|\to 0}\lambda_{\Delta, k}=\lambda_k \label{A4}.
	\end{equation}
\end{theorem}

\begin{remark} 
	Condition (\ref{A2}) implies that
	$\|\Psi_k(\lambda)-\Psi_k(\nu)\|\leq \eta_k |\lambda-\mu|, 
	\forall \lambda,\mu\in [0,\tau]$ (Lipschitz continuity).
    In particular,  $\Psi_k(\cdot)$ is strongly continuous, 
	so that the mapping $\Psi_k(\cdot):[0,t]\to\MSH$ 
	is a curve in $\MSH$.
\end{remark}

{\it Proof}. 
By using \cite[THEOREM 4.2]{AF}, the assumptions 
(\ref{A0})--(\ref{A3}) imply that
\begin{equation}
	\lim_{|\Delta |\to 0}\gamma_{\Delta, k}=1. \label{gamma}
\end{equation}
On the other hand, we can estimate $\epsilon_{\Delta, k}$ as follows.
\begin{eqnarray}
	\epsilon_{\Delta, k}
	&=&\sum_{l=0}^{N-1}\sum_{\stackrel{k_0, \cdots, k_{N-1}}
	{\forall i > l, k_i = k , k_l \neq k}}
	\lambda_{k_0}\prod _{j=1}^N\left|\langle \Psi_{k_j}(t_j), 
    e^{-i\Delta_j H} 
	\Psi_{k_{j-1}}(t_{j-1})\rangle \right|^2\\
	&=&\sum_{l=0}^{N-1}\left\{ \prod_{j=l+2}^{N}
	\left|\langle \Psi_{k}(t_j), e^{-i\Delta_j H} 
	\Psi_{k}(t_{j-1})\rangle \right|^2 
	\sum_{k_l, k_l \neq k}
	\left|\langle \Psi_{k}(t_{l+1}), e^{-i\Delta_{l+1} H} 
	\Psi_{k_l}(t_{l})\rangle \right|^2\right. \notag \\
	&&\left. \times \sum_{k_{l-1}}\left|\langle \Psi_{k_l}(t_l), 
    e^{-i\Delta_l H} \Psi_{k_{l-1}}(t_{l-1})\rangle \right|^2 
	\times \cdots \times 
	\sum_{k_{0}}\left|\langle \Psi_{k_1}(t_1), e^{-i\Delta_1 H} 
	\Psi_{k_{0}}(t_{0})\rangle \right|^2\lambda_{k_0}
	\right\}, \notag \\ 
	\label{epsilon}
\end{eqnarray}
in the case where $l=0, N-1$, $\{\ \cdots\}$ 
in (\ref{epsilon}) is given by
\begin{eqnarray}
	&&\prod_{j=2}^{N}
	\left|\langle \Psi_{k}(t_j), e^{-i\Delta_j H} 
	\Psi_{k}(t_{j-1})\rangle \right|^2 
	\sum_{k_0, k_0 \neq k}
	\left|\langle \Psi_{k_1}(t_1), e^{-i\Delta_1 H} 
	\Psi_{k_{0}}(t_{0})\rangle \right|^2\lambda_{k_0}, \\
	&&\sum_{k_{N-1}, k_{N-1} \neq k}
	\left|\langle \Psi_{k}(t_{l+1}), e^{-i\Delta_{l+1} H} 
	\Psi_{k_l}(t_{l})\rangle \right|^2
	\sum_{k_{N-2}}\left|\langle \Psi_{k_{N-1}}(t_{N-1}), 
    e^{-i\Delta_{N-1} H} 
	\Psi_{k_{l-2}}(t_{l-2})\rangle \right|^2 \notag \\
	&&\cdots 
	\times \sum_{k_{0}}\left|\langle \Psi_{k_1}(t_1), 
    e^{-i\Delta_1 H} 
	\Psi_{k_{0}}(t_{0})\rangle \right|^2\lambda_{k_0},
\end{eqnarray}
respectively.

By the Schwarz inequality, we have
\begin{eqnarray*}
	\prod_{j=l+2}^{N}
	\left|\langle \Psi_{k}(t_j), e^{-i\Delta_j H} 
	\Psi_{k}(t_{j-1})\rangle \right|^2 
	&\leq& \prod_{j=l+2}^{N}\|\Psi_{k}(t_j)\|^2\cdot \|
    e^{-i\Delta_j H} \Psi_{k}(t_{j-1})\|^2 \\
	&\leq& 1, \qquad \forall l \in \{0, \cdots ,N-2\}.
\end{eqnarray*}
For all $l\geq 1$,
\begin{eqnarray*}
&&\sum_{k_{l-1}}\left|\langle \Psi_{k_l}(t_l), e^{-i\Delta_l H} 
\Psi_{k_{l-1}}(t_{l-1})\rangle \right|^2 \times
\cdots 
\times \sum_{k_{0}}\left|\langle \Psi_{k_1}(t_1), e^{-i\Delta_1 H} 
\Psi_{k_{0}}(t_{0})\rangle \right|^2\lambda_{k_0}\\
&\leq& \sum_{k_{l-1}}\left|\langle \Psi_{k_l}(t_l), e^{-i\Delta_l H} 
\Psi_{k_{l-1}}(t_{l-1})\rangle \right|^2 \times
\cdots 
\times \sum_{k_{0}}\left|\langle e^{i\Delta_1 H} \Psi_{k_1}(t_1), 
\Psi_{k_{0}}(t_{0})\rangle \right|^2\\
&\leq&\sum_{k_{l-1}}\left|\langle \Psi_{k_l}(t_l), e^{-i\Delta_l H} 
\Psi_{k_{l-1}}(t_{l-1})\rangle \right|^2 \times
\cdots 
\times \| e^{i\Delta_1 H} \Psi_{k_1}(t_1)\|^2\\
&\leq&\dots \leq 1.
\end{eqnarray*}
Thus (\ref{epsilon}) implies that
\begin{equation}
	\epsilon_{\Delta, k}\leq 
\sum_{l=0}^{N-1}
 \sum_{k_l, k_l \neq k}
\left|\langle \Psi_{k}(t_{l+1}), e^{-i\Delta_{l+1} H} 
\Psi_{k_l}(t_{l})\rangle \right|^2. \label{hyouka215}
\end{equation}
In the case where $k_l\neq k$, we have $\lang \Psi_k(t_l), \Psi_{k_l}(t_l)\rang =0$.
Hence
\begin{eqnarray}
	&&\sum_{k_l, k_l \neq k}
\left|\langle \Psi_{k}(t_{l+1}), e^{-i\Delta_{l+1} H} 
\Psi_{k_l}(t_{l})\rangle \right|^2\notag \\
&=&
\sum_{k_l, k_l \neq k}
\left|
\langle \Psi_{k}(t_{l+1}), (e^{-i\Delta_{l+1} H}-1) \Psi_{k_l}(t_{l})\rangle 
+
\langle \Psi_{k}(t_{l+1})-\Psi_{k}(t_{l}), \Psi_{k_l}(t_{l})\rangle
\right|^2\notag \\
&\leq&
2\sum_{k_l, k_l \neq k}\left\{
\left|
\langle (e^{i\Delta_{l+1} H}-1)\Psi_{k}(t_{l+1}), \Psi_{k_l}(t_{l})\rangle 
\right|^2
+
\left|
\langle \Psi_{k}(t_{l+1})-\Psi_{k}(t_{l}), \Psi_{k_l}(t_{l})\rangle
\right|^2 \right\}\notag \\
&\leq& 2\left\{ \| (e^{i\Delta_{l+1} H}-1)\Psi_{k}(t_{l+1})\|^2
+ \| \Psi_{k}(t_{l+1})-\Psi_{k}(t_{l})\|^2
\right\}.\label{hyouka_o}
\end{eqnarray}
Let $E_H(\cdot)$ be the spectral measure of Hamiltonian $H$.
By the spectral theorem, we have
\begin{eqnarray}
\| (e^{i\Delta_{l+1} H}-1)\Psi_{k}(t_{l+1})\|^2
&=&\int_{\mathbb{R}}|e^{i\Delta_{l+1} x}-1|^2d\|E_H(x)\Psi_{k}(t_{l+1})\|^2
\notag \\
&\leq&\int_{\mathbb{R}}\Delta_{l+1}^2x^2d\|E_H(x)\Psi_{k}(t_{l+1})\|^2
\notag \\
&\leq&\Delta_{l+1}^2\|H\Psi_{k}(t_{l+1})\|^2.\label{hyouka_hami}
\end{eqnarray}

The assumptions (\ref{A0})--(\ref{A2}) imply that
\begin{equation}
\|H\Psi_{k}(t_{l+1})\|^2\leq \xi_k^2, \quad
\| \Psi_{k}(t_{l+1})-\Psi_{k}(t_{l})\|^2\leq \Delta_{l+1}^2\eta_k^2.
\label{hyouka_la}
\end{equation}
Therefore, (\ref{hyouka215}), (\ref{hyouka_o}), (\ref{hyouka_hami})
and (\ref{hyouka_la}) implies that
\begin{eqnarray}
\epsilon_{\Delta, k}
&\leq& 2\sum_{l=0}^{N-1}\left\{ \| (e^{i\Delta_{l+1} H}-1)\Psi_{k}(t_{l+1})\|^2
+ \| \Psi_{k}(t_{l+1})-\Psi_{k}(t_{l})\|^2
\right\}\notag \\
&\leq& 2\sum_{l=0}^{N-1}\left\{ \Delta_{l+1}^2
\|H\Psi_{k}(t_{l+1})\|^2+\| \Psi_{k}(t_{l+1})-\Psi_{k}(t_{l})\|^2\right\}\notag \\
&\leq &2(\xi_k^2+\eta_k^2)\sum_{l=1}^N\Delta_l^2. \label{hyoukaep}
\end{eqnarray}
By \cite[LEMMA 2.2]{AF}, 
$$
\lim_{|\Delta|\to 0}\sum_{l=1}^N\Delta_l^2=0.
$$
Thus (\ref{hyoukaep}) implies that
$\lim_{|\Delta |\to 0}\epsilon_{\Delta, k}=0$.
Hence, by 
(\ref{decomep}) and (\ref{gamma}), we obtain (\ref{A4})

\hfill \qed

\medskip

\begin{remark}
Assume that the conditions of Theorem \ref{saisho} hold.
Let $a>1$ be a constant and take $|\Delta |$ such that
\begin{equation}
(\xi_k^2+2\xi_k\eta_k)|\Delta |^2 + 2\eta_k |\Delta | \leq 
\frac{\log a}{a}. \label{lok}
\end{equation}
By the proof of \cite[THEOREM 4.2]{AF},
\begin{eqnarray}
&&\exp \left[-a \left\{(\xi_k^2+2\xi_k \eta_k)\sum_{l=1}^N\Delta_l^2 -2\sum_{l=1}^N\Re \lang \Psi_k(t_l)-\Psi_k(t_{l-1}), \Psi_k(t_{l-1})\rang\right\}\right] \notag \\
&&\leq \gamma_{\Delta , k}
\leq 1.\label{loga}
\end{eqnarray}
Then, by (\ref{hyoukaep}) and (\ref{loga}), we have
\begin{eqnarray}
&&| \lambda_{\Delta , k}-\lambda_k |
=|\lambda_k(\gamma_{\Delta , k}-1)+\epsilon_{\Delta, k}|
\leq \lambda_k(1-\gamma_{\Delta , k})+\epsilon_{\Delta, k}\notag \\
&\leq &\lambda_k\left( 1- 
\exp \left[-a \left\{(\xi_k^2+2\xi_k \eta_k)\sum_{l=1}^N\Delta_l^2 -2\sum_{l=1}^N\Re \lang \Psi_k(t_l)-\Psi_k(t_{l-1}), \Psi_k(t_{l-1})\rang\right\}\right]\right)\notag \\
&&+2(\xi_k^2+\eta_k^2)\sum_{l=1}^N\Delta_l^2.
\end{eqnarray}
\end{remark}

\medskip

The following corollary can be easily proven by using \cite[COROLLARY 4.4]{AF}.
\begin{corollary}\label{E1}
Assume that there exists $k\in \mathbb{N}$ such that the following conditions hold:
\begin{eqnarray}
&&\Psi_k(\cdot):[0,\tau]\to \mathcal{H} \quad
\text{is a strongly differentiable mapping}, \label{E11}\\
&&\forall \lambda \in [0,\tau], \quad \Psi_k(\lambda)\in D(H), \label{E12}\\
&&\xi_k < \infty , \label{E13}\\
&&\sup_{0\leq \lambda \leq \tau}\|\Psi_k'(\lambda) \|< \infty, \label{E14}\\
&&\text{where} \ \Psi_k'(\cdot) \ \text{denotes the strong derivative of} \
 \Psi_k(\cdot).\notag
\end{eqnarray}
Then (\ref{A0})--(\ref{A3}) hold.
Therefore, by Theorem \ref{saisho}, (\ref{A4}) holds.
\end{corollary}

\begin{example}\label{E2}
Let $A$ be a self-adjoint operator on $\mathcal{H}$.
Assume that there exists $k\in \mathbb{N}$ such that the following conditions hold:
\begin{eqnarray}
&&\Psi_k \in D(A) \cap \bigcap_{0\leq\lambda\leq \tau}D(He^{-i\lambda A}), 
\label{E21}\\
&&\sup_{0\leq\lambda\leq \tau}\|He^{-i\lambda A}\Psi_k\| < \infty , 
\label{E22}\\
&&\forall \lambda \in [o,\tau], \ \Psi_k(\lambda) = e^{-i\lambda A} \Psi_k .
\label{E23}
\end{eqnarray}
In this case, by \cite[EXAMPLE 4.5]{AF}, (\ref{E11})--(\ref{E14}) hold.
Then by using Corollary \ref{E1}, (\ref{A0})--(\ref{A4}) hold.
\end{example}

\medskip


\subsection{Trace norm convergence}

For the decomposition (\ref{schatten}), we define
	\begin{equation}
		\rho(t)
		:= \sum_n \lambda_n|\Psi_n(t)\rangle \langle \Psi_n(t)|,
		\quad \forall t \in [0,\tau].
	\end{equation}
Let us consider conditions of convergence from $\rho_{\Delta}(\tau)$
to $\rho(\tau)$ in the trace norm sense.

\begin{theorem}\label{nibannme}
Assume that the conditions (\ref{A0})--(\ref{A3}) hold
for all $k \in \mathbb{N}$ satisfying $\lambda_k >0$.

Then we have
\begin{eqnarray}
	\lim_{|\Delta |\to 0}\| \rho_{\Delta}(\tau)
- \rho(\tau)\|_1=0.\label{trace1}
\end{eqnarray}
\end{theorem}

{\it Proof}.
By definition of $\rho_{\Delta}(\tau)$, $\rho(\tau)$, and 
equation (\ref{decomep}), we have
\begin{eqnarray}
\| \rho_{\Delta}(\tau)
- \rho(\tau)\|_1
&=&\sum_k\langle \Psi_k(\tau), 
|\rho_{\Delta}(\tau)
- \rho(\tau)| \Psi_k(\tau)
\rangle \notag \\
&=&\sum_k|\lambda_{\Delta, k}-\lambda_k| \label{koyuuchi} \notag \\
&=&\sum_k|\lambda_k(\gamma_{\Delta,k}-1)+\epsilon_{\Delta,k}|\notag \\
&\leq&\sum_k\lambda_k(1-\gamma_{\Delta,k})
+\sum_k\epsilon_{\Delta,k}\notag \\
&=&\sum_k\lambda_k(1-\gamma_{\Delta,k})
+\sum_k(\lambda_{\Delta, k}-\lambda_k\gamma_{\Delta,k})\notag \\
&=&
2-2\sum_k\lambda_k\gamma_{\Delta,k}. \label{rholeq}
\end{eqnarray}
Note that
$$
|\lambda_k\gamma_{\Delta,k}|\leq \lambda_k \
(\forall k \in \mathbb{N}), \quad 
\sum_k\lambda_k=1.
$$
The assumptions (\ref{A0})--(\ref{A3}) imply that
$$\lim_{|\Delta|\to0}\lambda_k\gamma_{\Delta,k}=\lambda_k \
(\forall k \in \mathbb{N}).
$$
Hence, by using Lebesgue's dominated convergence theorem, we have
$$
\lim_{|\Delta|\to0}\sum_k\lambda_k\gamma_{\Delta,k}=1.
$$
Therefore, by (\ref{rholeq}), we obtain (\ref{trace1}).

\hfill \qed

\medskip

\begin{remark}
Assume that the conditions of Theorem \ref{nibannme} hold
and that $\sup_{k, \lambda_k\neq 0}\xi_k < \infty$
and $\sup_{k, \lambda_k\neq 0}\eta_k < \infty$ hold.
Then, for $a>1$, we can take $|\Delta |$ such that
(\ref{lok}) holds for all $k$ with $\lambda_k \neq 0$.
Then we have (\ref{loga}) for all $k\in \mathbb{N}$ 
with $\lambda_k \neq 0$. 
Hence, by (\ref{rholeq}), for all $k\in \mathbb{N}$, 
we obtain the following estimation:
\begin{eqnarray}
&&|\lambda_{\Delta , k}-\lambda_k | \leq \| \rho_{\Delta}(\tau)
- \rho(\tau)\|_1
\notag \\
&\leq& 2-2\sum_k\lambda_k
\exp \left[-a \left\{(\xi_k^2+2\xi_k \eta_k)\sum_{l=1}^N\Delta_l^2 -2\sum_{l=1}^N\Re \lang \Psi_k(t_l)-\Psi_k(t_{l-1}), \Psi_k(t_{l-1})\rang\right\}\right].\notag \\
\end{eqnarray}
\end{remark}

\medskip

The following corollary and example can be easily proven by using 
Corollary \ref{E1}, Example \ref{E2}, and Theorem \ref{nibannme}.

\begin{corollary}\label{c3}
Assume that the conditions (\ref{E11})--(\ref{E14}) hold 
for all $k \in \mathbb{N}$ with $\lambda_k >0$.
Then we have (\ref{trace1}).
\end{corollary}

\begin{example}\label{c4}
Let $A$ be a self-adjoint operator on $\mathcal{H}$.
Assume that the conditions (\ref{E21})--(\ref{E23}) hold 
for all $k \in \mathbb{N}$ with $\lambda_k >0$.
Then we have (\ref{trace1}).
\end{example}

In Example \ref{c4}, let us consider the case of $d<\infty$.
It is easy to see that 
the assumptions (\ref{E21})--(\ref{E22}) are satisfied.
On the other hand, by Stone's theorem, for all $U \in \mathfrak{U}(\mathcal{H})$, 
there exists a self-adjoint operator $A$ such that $U = e^{-i\tau A}$. 
Since $\rho(\tau)= U\rho U^*$, we have 
$
\lim_{|\Delta |\to 0}\| \rho_{\Delta}(\tau)
- U\rho U^*\|_1=0.
$
This fact shows that, in the case $d<\infty$, 
an arbitrary state in 
$\{U\rho U^*\ | \ U \in \mathfrak{U}(\mathcal{H}) \}$
can be approximated (in the trace norm sense) 
by states obtained after an appropriate continuous measurements.
In other words, in this case, we can approximate 
an arbitrary unitary channel
by continuous quantum measurements.

\subsection{Application to quantum Zeno effect for mixed states}

Let $\Psi_k \in D(H)$ and 
$\Psi_k(\lambda)=\Psi_k \ (\forall \lambda \in [0,\tau])$
holds for all $k\in \mathbb{N}$ with $\lambda_k >0$.

This is the case where $A=0$ in Example \ref{c4}.
Then (\ref{A0})--(\ref{A3}) hold for all $k\in \mathbb{N}$
with $\lambda_k >0$.
Hence, we have (\ref{trace1}).

This means that, by the series of measurement with respect to 
the family of the projection
operators $\{|\Psi_k \rangle \langle \Psi_k | \}_k$,
transitions to states different from the initial state are hindered.
This can be interpreted as a quantum Zeno effect for 
mixed states.


\section{Convergence condition of the von Neumann entropy}

Let $\varphi:[0,\infty )\ni \lambda \mapsto -\lambda \log \lambda \in [0,\infty)$,
where $\varphi (0):=0$.
Then $\varphi$ is continuous, concave, and subadditive.
Let 
$S(\rho)$ be the von Neumann entropy of $\rho\in\mathfrak{S}(\mathcal{H})$.
i.e. 
$$
S(\rho):= \mathrm{Tr}\varphi(\rho).
$$

In the case $d < \infty$, 
by Fannes' inequality, 
we have for all $\rho_1, \rho_2 \in \mathfrak{S}(\mathcal{H})$
$$
\|\rho_1 -\rho_2 \|_1 \leq 1/e
\Longrightarrow
|S(\rho_1)-S(\rho_2)|\leq \|\rho_1 -\rho_2 \|_1\log d
+\varphi(\|\rho_1 -\rho_2 \|_1).
$$
Therefore the von Neumann entropy is continuous with respect to the trace norm.

On the other hand，in the case $d= \infty$，
although the von Neumann entropy is lower semi-continuous 
with respect to the trace norm
（i.e. $\lim_{n\to \infty}\|\rho_n -\rho \|_1 = 0
\Rightarrow S(\rho)\leq \liminf_{n\to \infty}S(\rho_n)$),
it is not necessarily continuous.
Moreover, it is known that the set 
$\{ \rho \in \mathfrak{S}(\mathcal{H})\ | \ S(\rho) < \infty \}$
is of the first category \cite{We}.

In what follows, we deal with the case where $d = \infty$ only.

For $\rho_{\Delta}(\tau)$ and $\rho$ considered in the section 2, 
conditions of the convergence 
$S(\rho_{\Delta}(\tau))\to S(\rho)$ are given by 
the following theorem.

\begin{theorem}\label{vNentropy}
Assume that the conditions (\ref{A0})--(\ref{A2}) hold 
for all $k \in \mathbb{N}$, and that the condition (\ref{A3})
holds for all $k \in \mathbb{N}$ with $\lambda_k >0$.
Suppose that the following conditions hold:
\begin{eqnarray}
&&\xi_k \to 0, \quad \eta_k \to 0 \ (k\to 0), \label{main1}\\
&&S(\rho)< \infty, \label{main2}\\
&&\sum_k \varphi (\xi_k^2) <\infty , \quad
\sum_k \varphi (\eta_k^2) <\infty. \label{main3}
\end{eqnarray}
Then
\begin{eqnarray}
	\lim_{|\Delta |\to 0}S(\rho_{\Delta}(\tau))=S(\rho (\tau))=S(\rho).
\end{eqnarray}
\end{theorem}

\begin{remark}\label{vNrem1}

The function $\varphi$ is monotone increasing on $[0,1/e]$ and
$$
\xi_k^2=\sup_{0\leq \lambda \leq \tau}\| H\Psi_k(\lambda )\|^2
=\sup_{0\leq \lambda \leq \tau}\int_{\mathbb{R}}x^2d\|E_H(x)\Psi_k(\lambda)\|^2.
$$
Hence, $\xi_k\to0\ (k\to \infty)$ implies
that there exists $N_0 \in \mathbb{N}$ such that, for all $k>N_0$,
$$
\varphi(\xi_k^2)\geq \sup_{0\leq \lambda \leq \tau}
\varphi \left( \int_{\mathbb{R}}x^2d\|E_H(x)\Psi_k(\lambda)\|^2 \right).
$$
By Jensen's inequality, we have
$$
\varphi \left( \int_{\mathbb{R}}x^2d\|E_H(x)\Psi_k(\lambda)\|^2 \right)
\geq
\int_{\mathbb{R}}\varphi(x^2)d\|E_H(x)\Psi_k(\lambda)\|^2.
$$
Hence, for all $k> N_0$,
$$
\varphi(\xi_k^2) 
\geq
\sup_{0\leq \lambda \leq \tau}
\int_{\mathbb{R}}\varphi(x^2)d\| E_H(x)\Psi_k(\lambda)\|^2.
$$
 Then, we have
$$
\forall k>N_0, \ \forall \lambda \in [0,\tau], \
\Psi_k(\lambda)\in D(\sqrt{\varphi(H^2)}), \ 
\varphi(\xi_k^2)\geq \sup_{0\leq \lambda \leq \tau}
\|\sqrt{\varphi(H^2)}\Psi_k(\lambda) \|^2.
$$
Moreover, using the estimate that
\begin{eqnarray*}
\sum_k\varphi(\xi_k^2)=\sum_{k=1}^{N_0}\varphi(\xi_k^2)
+\sum_{k=N_0+1}^{\infty}\varphi(\xi_k^2)
\geq 
\sum_{k=1}^{N_0}\varphi(\xi_k^2)
+
\sum_{k=N_0+1}^{\infty}
\sup_{0\leq \lambda \leq \tau}
\|\sqrt{\varphi(H^2)}\Psi_k(\lambda) \|^2 \\
\geq 
\sum_{k=1}^{N_0}\varphi(\xi_k^2)
+
\sup_{0\leq \lambda \leq \tau}\sum_{k=N_0+1}^{\infty}
\|\sqrt{\varphi(H^2)}\Psi_k(\lambda) \|^2,
\end{eqnarray*}
we obtain
\begin{equation}
\xi_k\to 0 \ (k\to \infty), \ \sum_k\varphi(\xi_k^2)<\infty
\Longrightarrow
\exists N_0 \in \mathbb{N}, \ 
\sup_{0\leq \lambda \leq \tau}\sum_{k=N_0+1}^{\infty}
\|\sqrt{\varphi(H^2)}\Psi_k(\lambda) \|^2 < \infty.\label{varphi}
\end{equation}

Particularly, in the case $H\in \mathfrak{B}(\MSH)$,
we have, for all $\Phi \in \mathcal{H}$,
$$
\int_{\mathbb{R}}\varphi(x^2)d\|E_H(x)\Phi\|^2
\leq 
\sup_{x\in \sigma(H)}\varphi(x^2)\int_{\mathbb{R}}d\|E_H(x)\Phi \|^2
=\sup_{x\in \sigma(H)}\varphi(x^2)\cdot \|\Phi \| < \infty.
$$
Hence, we obtain $\sqrt{\varphi(H^2)}\in \mathfrak{B}(\MSH)$.
Therefore, by (\ref{varphi}), we have 
\begin{equation}
\xi_k\to 0 \ (k\to \infty), \ \sum_k\varphi(\xi_k^2)<\infty
\Longrightarrow
\varphi(H^2)\in \mathfrak{T}(\MSH).
\end{equation}
We remark that, in this case, if Hamiltonian $H$ is 
represented as a density operator, then 
$\varphi(H^2)\in \mathfrak{T}(\MSH)$ means $S(H^2)<\infty $.
\end{remark}

{\it Proof}.
The assumption of this theorem and Theorem \ref{nibannme} imply that 
$
	\lim_{|\Delta |\to 0}\| \rho_{\Delta}(\tau)
- \rho(\tau)\|_1=0.
$
Hence we have 
$\text{w}\mbox{-}\lim_{|\Delta |\to 0}\rho_{\Delta}(\tau)=\rho (\tau)$, 
where $\text{w}\mbox{-}\lim$ means weak limit.

By 
(\ref{decomep}), (\ref{hyoukaep}) and $\gamma_{\Delta , k}\leq 1$, we have
$$
\lambda_{\Delta, k}\leq \lambda_{k}+2(\xi_k^2+\eta_k^2)\sum_{l=1}^{N}\Delta_l^2.
$$
By $\lim_{|\Delta |\to 0}\sum_{l=1}^{N}\Delta_l^2=0$, 
there exists $\delta >0$ such that, for all $\Delta$,
$
|\Delta |<\delta \Rightarrow \sum_{l=1}^{N}\Delta_l^2
<1/2.
$
Thus
\begin{equation}
\lambda_{\Delta, k}\leq \lambda_{k}+\xi_k^2+\eta_k^2 \quad
(|\Delta |<\delta).\label{lren}
\end{equation}
We set
\begin{eqnarray}
\sigma := \sum_k(\lambda_k + \xi_k^2 +\eta_k^2)|\Psi_k(\tau)\rangle \langle
\Psi_k(\tau)|.
\end{eqnarray}
By the assumption of this theorem,
$\sigma \in \mathfrak{C}(\MSH)$. 
On the other hand, (\ref{lren}) implies that
\begin{equation}
\rho_{\Delta}(\tau)\leq \sigma \quad (|\Delta |<\delta).
\end{equation}
Moreover, 
by the assumption of this theorem and subadditivity of $\varphi$, 
we have
\begin{eqnarray}
S(\sigma)&=&\sum_k \varphi(\lambda_k +\xi_k^2+\eta_k^2)\\
&\leq& S(\rho)+\sum_k\varphi(\xi_k^2)+\sum_k\varphi(\eta_k^2)<\infty.
\end{eqnarray}
Hence, by Simon's dominated convergence theorem
for entropy \cite[THEOREM A.3]{LR}, 
we have
$$
	\lim_{|\Delta |\to 0}S(\rho_{\Delta}(\tau))=S(\rho (\tau)).
$$
It is obvious that 
$S(\rho (\tau))=S(\rho)$ holds.
\hfill \qed

\begin{remark}
In the proof of Theorem \ref{vNentropy}, 
we used that 
\begin{equation}
S(\rho)<\infty, \ \sum_k\varphi(\xi_k^2)<\infty, \ \sum_k\varphi(\eta_k^2)<\infty 
\Longrightarrow
\sum_k \varphi(\lambda_k +\xi_k^2+\eta_k^2) <\infty.
\end{equation}
Conversely, we can show that, under condition (\ref{main1}),
\begin{equation}
\sum_k \varphi(\lambda_k +\xi_k^2+\eta_k^2) <\infty
\Longrightarrow
S(\rho), \ \sum_k\varphi(\xi_k^2), \ \sum_k\varphi(\eta_k^2)<\infty
\label{follow}
\end{equation}
as follows.
By 
$\lambda_k+\xi_k^2+\eta_k^2 \to 0 \ (k\to \infty)$, 
we have
$$
\exists N_0 \in \mathbb{N}, 
\forall k > N_0, 
\max \{\lambda_k, \xi_k^2, \eta_k^2 \}\leq \lambda_k+\xi_k^2+\eta_k^2 < 1/e.
$$
Hence, by the fact that $\varphi$ is a monotone increasing function on $[0,1/e]$,
we obtain
$$
\max
\left\{
\sum_{k=N_0+1}^{\infty}\varphi(\lambda_k), 
\sum_{k=N_0+1}^{\infty}\varphi(\xi_k^2), 
\sum_{k=N_0+1}^{\infty}\varphi(\eta_k^2)
\right\}
\leq
\sum_{k=N_0+1}^{\infty}\varphi(\lambda_k + \xi_k^2+ \eta_k^2).
$$
Therefore, we have (\ref{follow}).
Thus, in Theorem \ref{vNentropy}, 
we can replace the condition (\ref{main2}) and (\ref{main3})
with $\sum_k \varphi(\lambda_k +\xi_k^2+\eta_k^2) <\infty $.
\end{remark}

\begin{example}\label{saigo}
Let $A$ be a self-adjoint operator on $\mathcal{H}$.
Assume that
$A, H \in \mathfrak{C}(\MSH)$, and that
$A$ and $H$ are strongly commuting.
Moreover, we assume that
\begin{eqnarray}
&&\forall k \in \mathbb{N}, \ 
\forall \lambda \in [0, \tau], \ \Psi_k(\lambda)=e^{-i\lambda A}\Psi_k ,
\label{vNe1}\\
&&S(\rho)< \infty , \
\sum_{k}\varphi(\| H\Psi_k \|^2)< \infty , \
\sum_{k}\varphi(\| A\Psi_k \|^2)< \infty . \label{vNe2}
\end{eqnarray}
Then, the compactness, the strong commutativity 
of $A$ and $H$, and (\ref{vNe1}) imply that
$\xi_k=\| H\Psi_k \| \to 0 , \
\eta_k=\| A\Psi_k \| \to 0 \ (k\to \infty)$.
Hence, the assumption of Theorem \ref{vNentropy} is satisfied.
Hence, we have 
$S(\rho_{\Delta}(\tau))
\to S(\rho)\ (|\Delta |\to 0)$.
\end{example}

In Example \ref{saigo}, let us consider the case of $A=0$.
The following fact can be easily seen:
\begin{eqnarray}
&&H \in \mathfrak{C}(\MSH), \
\Psi_k(\lambda)=\Psi_k \ 
(\forall k \in \mathbb{N}, \ 
\forall \lambda \in [0, \tau]), \
S(\rho)< \infty , \ \sum_{k}\varphi(\| H\Psi_k \|^2)< \infty \notag \\
&&\Longrightarrow 
\lim_{|\Delta |\to 0}S(\rho_{\Delta}(\tau)) = S(\rho).\label{compact1}
\end{eqnarray}
This is the case of QZE.
We remark that, if $\{\Psi_k \}_k$ is a sequence
 of eigenvectors of $H$, we have 
$\sum_{k}\varphi(\| H\Psi_k \|^2)=\mathrm{Tr}\varphi(H^2)<\infty$.
Then, in (\ref{compact1}), 
we can replace the condition $\sum_{k}\varphi(\| H\Psi_k \|^2)<\infty$
with $\varphi (H^2)\in \mathfrak{T}(\mathcal{H})$.


\section*{Acknowledgments}

The author would like to thank 
Professor Asao Arai for valuable comments.

\end{document}